\documentclass[conference]{IEEEtran}
\IEEEoverridecommandlockouts
\usepackage{cite}
\usepackage{hyperref}
\usepackage{amsmath,amssymb,amsfonts}
\usepackage{graphicx}
\usepackage{textcomp}
\usepackage{xcolor}
\usepackage{tikz}
\usepackage{caption}
\usepackage{graphicx}   
\usepackage{subcaption} 
\usepackage{multirow}   
\usepackage{booktabs} 
\usepackage{algorithm}
\usepackage{algpseudocode}
\usepackage{tabularx}
\usetikzlibrary{positioning, shapes, fpu}

\tikzset{
    declare function={
        gaussian(\x,\y,\z) = exp(-((\x-\y)/\z)^2/2);
        base(\x) = 1.2 + 0.3*cos(2*\x);
        lobes(\x) = 1.5*gaussian(\x, 90, 15) + 0.5*gaussian(\x, 210, 20) + 0.4*gaussian(\x, 330, 20);
    },
    plot_style/.style={very thick, blue!60!black, smooth},
    box_style/.style={draw, thick, rectangle, minimum height=3.5cm, minimum width=3.5cm},
    label_style/.style={text width=4cm, align=center},
}

\def\BibTeX{{\rm B\kern-.05em{\sc i\kern-.025em b}\kern-.08em
    T\kern-.1667em\lower.7ex\hbox{E}\kern-.125emX}}
\begin{document}

\title{SPARK: Sparse Parametric Antenna Representation using Kernels

\thanks{This research was supported in part by National Science Foundation (NSF) grant OAC-2512931.

© 2026 IEEE.  Personal use of this material is permitted.  Permission from IEEE must be obtained for all other uses, in any current or future media, including reprinting/republishing this material for advertising or promotional purposes, creating new collective works, for resale or redistribution to servers or lists, or reuse of any copyrighted component of this work in other works.} 

}

\author{
\IEEEauthorblockN{William Bjorndahl, Mark O'Hair, Ben Zoghi, and Joseph Camp}
\IEEEauthorblockA{Department of Electrical and Computer Engineering, Southern Methodist University\\
\{wbjorndahl, mohair, bzoghi, camp\}@smu.edu}
}

\maketitle

\begin{abstract}

Channel state information (CSI) acquisition and feedback overhead grows with the number of antennas, users, and reported subbands. This growth becomes a bottleneck for many antenna and reconfigurable intelligent surface (RIS) systems as arrays and user densities scale. Practical CSI feedback and beam management rely on codebooks, where beams are selected via indices rather than explicitly transmitting radiation patterns. Hardware-aware operation requires an explicit representation of the measured antenna/RIS response, yet high-fidelity measured patterns are high-dimensional and costly to handle. We present SPARK (Sparse Parametric Antenna Representation using Kernels), a training-free compression model that decomposes patterns into a smooth global base and sparse localized lobes. For 3D patterns, SPARK uses low-order spherical harmonics for global directivity and anisotropic Gaussian kernels for localized features. For RIS 1D azimuth cuts, it uses a Fourier-series base with 1D Gaussians. On patterns from the AERPAW testbed and a public RIS dataset, SPARK achieves up to 2.8$\times$ and 10.4$\times$ reductions in reconstruction MSE over baselines, respectively. Simulation shows that amortizing a compact pattern description and reporting sparse path descriptors can produce 12.65\% mean uplink goodput gain under a fixed uplink budget. Overall, SPARK turns dense patterns into compact, parametric models for scalable, hardware-aware beam management.

\end{abstract}


\section{Introduction}

Massive multiple-input multiple-output (MIMO) arrays and reconfigurable intelligent surfaces (RIS) enable highly directive transmission and programmable manipulation of the propagation environment. However, as array sizes and user densities grow, control-plane overhead increases. Specifically, channel state information (CSI) acquisition and reporting incur overhead, since CSI is required to perform beamforming and scheduling. In frequency-division duplex settings, pilot/reference signal overhead and feedback payload can grow rapidly with the number of antennas, users, and frequency resources. This growth motivates compression and overhead reduction mechanisms~\cite{etsi_ts_138214_v15.4.0_2019, boloursaz2020deep}.

CSI feedback and beam management typically rely on finite codebooks and structured reports. These include wideband and subband quantities such as channel quality indicator (CQI) and precoding matrix indicator (PMI), whose payload and configuration depend on the system design~\cite{etsi_ts_138214_v15.4.0_2019, mathworks5gNrDownlinkCsiReporting}. This codebook-based abstraction implicitly assumes an underlying set of beam shapes, which are ultimately determined by the antenna array or RIS hardware response.

An important component in this pipeline is the hardware radiation pattern, the directional gain response of the antenna or RIS configuration. While much analytical work assumes idealized array responses, practical hardware exhibits non-ideal main lobes, sidelobes, and nulls that vary with frequency and configuration~\cite{hassouna}. High-fidelity measured patterns are inherently high-dimensional (e.g., thousands of angular samples), making direct storage, distribution, or reuse of full pattern grids impractical at scale. This is especially relevant in emerging disaggregated and programmable RAN architectures, where multiple network functions and controllers may need a consistent, portable description of the hardware beam/pattern response~\cite{etsi_ts_103982_v8.0.0_2024, oran_architecture_overview_2025}.

These observations create a systems-level gap. Current approaches either (i) use coarse finite codebooks that reduce overhead but limit resolution and introduce mismatch, or (ii) assume access to high-fidelity hardware responses that are too expensive to store and share directly. This work targets the missing piece in this trade-off. We explore using a compact and interpretable representation of measured antenna/RIS patterns that can be stored and transmitted once (or infrequently). As a result, recurring overhead is shifted toward tracking the time-varying channel/environment rather than repeatedly paying for the static hardware response~\cite{8056991, 9435015}. This aligns with the goal of reducing CSI acquisition/feedback overhead in large-scale systems~\cite{ma2025low}.

We introduce SPARK (Sparse Parametric Antenna Representation using Kernels), a hybrid parametric model that decomposes a measured radiation pattern into complementary components: (i) a smooth global base and (ii) sparse localized features. For 3D antenna patterns, the global base is represented with a low-order spherical harmonic (SH) expansion, while localized lobes are modeled with a small number of anisotropic Gaussian kernels on the angular domain. For RIS measurements represented as 1D azimuthal cuts, we use a low-order Fourier series as the global base and 1D Gaussians for localized peaks. SPARK is training-free and produces physically meaningful parameters (e.g., lobe locations and widths), supporting both compression and semantic manipulation of patterns.

This representation enables a useful systems-level separation. The static, hardware-specific radiation response can be stored and communicated once in a compact form. Subsequent feedback can focus on sparse environmental descriptors such as dominant path directions. In addition, SPARK provides a continuous, high-resolution pattern-space virtual codebook. Rather than selecting only from a fixed finite set of beams, the network can evaluate the fitted radiation pattern at arbitrary angles and use its interpretable parameters for control. This is relevant for open and programmable RAN architectures like O-RAN, where near-real-time controllers can benefit from compact, semantically meaningful primitives for configuring beams and surfaces.

Overall, SPARK reduces overhead by transforming dense radiation pattern data into a sparse parametric model that is compact to store, fast to evaluate, and interpretable. The contributions of this work are:
\begin{enumerate}
    \item \textbf{Hybrid parametric representation:} We propose SPARK, a global-plus-local decomposition for antenna and RIS radiation pattern compression (SH/Fourier base + sparse Gaussian kernels).
    \item \textbf{Robust fitting procedure:} We develop a stable alternating refinement algorithm with prominence-based lobe initialization and bounded nonlinear optimization, producing reliable fits on measured pattern data without training.
    \item \textbf{Measured validation on two datasets:} We evaluate SPARK on (i) 3D wideband antenna patterns from the AERPAW anechoic chamber and (ii) a public RIS dataset of 1D azimuth cuts, demonstrating up to 2.8$\times$ and 10.4$\times$ reductions in reconstruction MSE over strong baselines, respectively~\cite{mma2-0t93-23, rossanese2022designing}.
    \item \textbf{System-level impact via simulation:} We quantify how reduced feedback payload can translate to uplink resource savings, showing a 12.65\% mean uplink goodput gain at 50 users under a fixed uplink budget in Monte Carlo simulation (isolating feedback payload overhead).
\end{enumerate}

\section{Related Work}
\label{sec:relatedworks}

\begin{figure*}[t]
    \centering
    \begin{tikzpicture}

\node[box_style,
    path picture={
        \foreach \r in {1,2} {\draw[gray!40, dashed] (path picture bounding box.center) circle (\r);}
        \draw[plot_style, shift={(path picture bounding box.center)}]
            plot[smooth cycle] coordinates { (90:2.7) (70:0.9) (50:1.1) (30:1.4) (0:1.5) (-30:1.4) (-50:1.1) (-70:0.9) (270:0.9) (250:1.1) (230:1.2) (210:1.7) (190:1.2) (170:1.1) (150:1.4) (130:1.1) (110:0.9) };
    }
] (original) {};
\node[label_style, below=0.1cm of original] {
    \small
    \textbf{Original Radiation Pattern} \\ (Complex and Dense)
};

\node[right=0.5cm of original] (equals) {\Huge $=$};

\node[box_style, right=0.5cm of equals,
    path picture={
        \foreach \r in {1,2} {\draw[gray!40, dashed] (path picture bounding box.center) circle (\r);}
        \draw[plot_style, shift={(path picture bounding box.center)}]
            plot[smooth cycle] coordinates { (90:0.9) (45:1.2) (0:1.5) (-45:1.2) (270:0.9) (225:1.2) (180:1.5) (135:1.2) };
    }
] (base) {};
\node[label_style, below=0.1cm of base] {
    \small
    \textbf{Global Base} \\ (Low-Order SH / Fourier)
};

\node[right=0.5cm of base] (plus) {\Huge $+$};

\node[box_style, right=0.5cm of plus,
    path picture={
        \foreach \r in {1,2} {\draw[gray!40, dashed] (path picture bounding box.center) circle (\r);}
        \draw[plot_style, shift={(path picture bounding box.center)}]
            plot[smooth] coordinates { (60:0) (75:0.2) (90:1.5) (105:0.2) (120:0) }
            plot[smooth] coordinates { (180:0) (195:0.2) (210:0.8) (225:0.2) (240:0) }
            plot[smooth] coordinates { (300:0) (315:0.2) (330:0.6) (345:0.2) (360:0) };
    }
] (local) {};
\node[label_style, below=0.1cm of local] {
    \small
    \textbf{Local Features} \\ (Sparse Gaussians)
};

\end{tikzpicture}
    \caption{Conceptual illustration of SPARK. A measured radiation pattern is represented as the sum of a smooth global base (low-order spherical harmonics) and a sparse set of localized lobes (anisotropic Gaussian kernels) that capture sharp features.}
    \label{fig:concept_diagram}
\end{figure*}
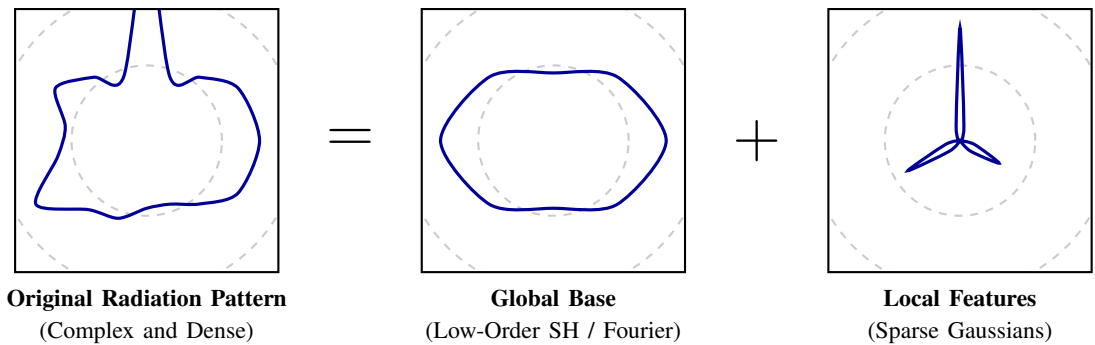

Efficient representations of antenna and RIS radiation patterns are relevant across antenna metrology, channel modeling, propagation simulation, and CSI feedback design. SPARK sits between two well-established modeling philosophies: (i) global expansions (e.g., spherical harmonics or Fourier-type bases) that compactly capture smooth structure, and (ii) localized primitives that efficiently represent narrow lobes, sidelobes, and deep nulls. We review related work in these areas and clarify SPARK's position.

\subsection{Antenna Metrology and Global-Basis Pattern Models}
Spherical near-field measurement and post-processing provide the classical foundation for representing 3D antenna patterns. Measured fields are expanded in spherical wave functions and transformed to far-field quantities that can be rotated, interpolated, and analyzed in a mathematically well-posed way~\cite{doi:10.1049/PBEW026E, IEEE1720-2012}. 

Within this paradigm, SHs are a common global basis for compact modeling of radiation patterns. Prior work has demonstrated SH-based pattern modeling for measured data, including polarimetric formulations and least-squares coefficient estimation, as well as practical operations such as rotation and interpolation~\cite{5068319,6175298,rahola2008processing}. These global-basis approaches are especially effective when the pattern varies smoothly over angle, since low-order coefficients capture broad directivity with few parameters.

A closely related representation used in geometry-based channel modeling is the effective aperture distribution function (EADF), which expresses polarimetric array responses via a spatial-frequency (discrete Fourier transform/Fourier) description that supports efficient interpolation and array-manifold derivatives~\cite{1394809, eadf}. Both EADF and SH produce a continuous pattern model from a finite set of coefficients. However, like other global bases, representing highly localized structure (very narrow main beams, rich sidelobe structure, sharp nulls) can require substantially higher orders, reducing compression gains and increasing evaluation cost.

\subsection{Localized and Sparse Directional Bases}
A complementary direction is to use localized bases on the sphere. Multiresolution constructions such as spherical wavelets provide localized basis functions that can represent localized features more efficiently than global harmonics~\cite{10.1145/218380.218439}. While this literature originates largely in graphics/numerical analysis, its insight is directly relevant to radiation patterns as spherical functions. Specifically, providing local support improves representation of localized structure.

Another widely used family of localized spherical primitives is spherical Gaussians and their anisotropic variants. These primitives compactly represent lobe-like structure with a small number of parameters and are commonly used to approximate anisotropic spherical functions~\cite{10.1145/2508363.2508386}. SPARK follows this same modeling intuition but in a deliberately lightweight form. Instead of using a full multiresolution transform, it places a small number of localized kernels at prominent residual peaks after a low-order global fit. This produces a compact and interpretable parameterization (lobe locations and widths) and supports stable fitting via bounded optimization and alternating refinement.

\subsection{CSI Feedback Compression and Overhead Reduction}
CSI acquisition and feedback has a large literature in massive MIMO. Practical systems typically rely on finite codebooks and structured CSI reporting formats (e.g., PMI/CQI), which reduce signaling but introduce quantization and resolution limits~\cite{etsi_ts_138214_v15.4.0_2019, 10.1109/JSAC.2008.081002}. Beyond codebooks, compressive sensing techniques exploit sparsity or low-dimensional structure in angular/delay domains to reduce estimation and feedback overhead, including spatially common sparsity across subcarriers and adaptive acquisition strategies~\cite{10.1109/TSP.2015.2463260}.

More recently, learning-based approaches use autoencoders and attention mechanisms to improve CSI rate--distortion performance, including CsiNet and successors~\cite{wen2018deep,CHEN2024102516}. Model-driven or hybrid learning approaches further combine signal structure with learnable components to reduce overhead and improve robustness~\cite{guo2021csi}. Recent work also explores separating quasi-static versus dynamic CSI components to reduce recurring feedback payload~\cite{10757893}. Surveys summarize the broader landscape of CSI acquisition/feedback in massive MIMO~\cite{boloursaz2020deep}.

SPARK is complementary to CSI-focused compression methods. Instead of compressing instantaneous CSI, it targets the static, hardware-specific radiation pattern that is often idealized or assumed known in system studies. By compressing measured patterns into a small set of interpretable parameters, SPARK supports pattern-aware schemes where recurring reporting can emphasize sparse environmental descriptors (e.g., dominant path directions), while the static pattern model is amortized over time.

\subsection{RIS Modeling and Measurement-Driven Characterization}
RIS research often uses simplified analytical models (e.g., idealized array responses) to enable tractable optimization and network-level analysis~\cite{hassouna}. At the same time, multiple works emphasize that practical RIS behavior can deviate from ideal models and that measurement-driven characterization is important for realistic evaluation and deployment~\cite{bjornson2020reconfigurable, ZHENG2023603, 9206044, ris-aided-wireless}. Efficient modeling that blends electromagnetic characterization with propagation simulation has also been studied, e.g., combining full-wave extraction of RIS scattering properties with ray-tracing style propagation~\cite{9833252}.

Our RIS study leverages a public measurement dataset from a switch-based RIS prototype~\cite{rossanese2022designing}, consisting of 1D azimuth cuts of received power for many RIS configurations. SPARK adapts naturally to this setting by using a low-order Fourier series as the global base and sparse 1D Gaussian kernels for localized peaks. This results in a compact and interpretable description of measured RIS behavior.

\subsection{Programmable RAN Context}
Finally, SPARK is motivated by emerging programmable RAN architectures that separate control logic from the underlying radio stack. In O-RAN, near-real-time control applications can benefit from compact, semantically meaningful abstractions of hardware and channel behavior~\cite{etsi_ts_103982_v8.0.0_2024}. SPARK's parametric form provides such an abstraction by exposing interpretable pattern primitives (e.g., lobe location and width) instead of opaque latent variables.

\section{Hybrid Spherical Harmonic-Gaussian Model}
\label{sec:model}

We present SPARK, a hybrid parametric representation for antenna radiation patterns. The idea is to decompose a measured pattern into (i) a smooth global component captured by a low-order SH expansion and (ii) a small number of localized kernels that capture sharp lobes and sidelobes. The resulting model is compact, interpretable, and efficient to evaluate once fitted.

\subsection{Theoretical Foundation}

\subsubsection{Data Preprocessing and Normalization}
Measured radiation patterns are commonly reported in logarithmic units (e.g., dB/dBm) on an angular grid. SPARK fits in a normalized linear (power-like) domain. Concretely, given log-scale samples $P_{\log}(\cdot)$, we first convert to linear power
\begin{equation}
P_{\mathrm{lin}}(\cdot) = 10^{P_{\log}(\cdot)/10},
\end{equation}
and then apply per-pattern min--max normalization
\begin{equation}
G(\cdot)=\frac{P_{\mathrm{lin}}(\cdot)-\min P_{\mathrm{lin}}}{\max P_{\mathrm{lin}}-\min P_{\mathrm{lin}}+\epsilon}\in[0,1],
\label{eq:minmax_norm}
\end{equation}
with a small $\epsilon>0$ for numerical stability. This normalization preserves the directional shape of the pattern (relative lobes and nulls) while removing the absolute gain scale. If absolute calibration is needed, it can be retained separately as a scalar (e.g., the original peak value, or the min/max values) alongside the compressed SPARK parameters.

Fitting in the normalized linear domain is important for two reasons. First, SPARK uses an additive decomposition $G \approx G_{\text{base}} + G_{\text{local}}$, which produces a linear least-squares update for the global basis coefficients and a bounded nonlinear least-squares update for the local kernels. Fitting directly in dB would introduce a logarithmic nonlinearity that couples the global and local components and changes the implied error weighting. Second, least-squares error in dB can overweight low-power regions (e.g., deep nulls) where measurement noise may dominate. Fitting in linear power provides a more stable balance between main-lobe structure and sidelobe/null fidelity under our bounded, alternating refinement procedure.

\subsubsection{Antenna Pattern Structure}
We consider a measured (normalized) far-field gain pattern $G(\theta,\phi)$ on the sphere, where $\theta\in[0,\pi]$ denotes the polar angle and $\phi\in[0,2\pi)$ denotes azimuth. Measured patterns can exhibit a separation of scales. For example, a slowly varying background shaped by element/ground-plane effects, and localized directive features produced by interference and array effects. We model the pattern as
\begin{equation}
G(\theta,\phi) \;=\; G_{\text{base}}(\theta,\phi) \;+\; G_{\text{local}}(\theta,\phi) \;+\; \epsilon(\theta,\phi),
\end{equation}
where $G_{\text{base}}$ captures smooth global structure, $G_{\text{local}}$ captures localized lobes and null structure, and $\epsilon$ captures measurement noise and unresolved fine-scale variations. A conceptual illustration showing the composition of the global and local features is shown in Fig.~\ref{fig:concept_diagram}.

\subsubsection{Basis Selection}
\textbf{Spherical harmonics for $G_{\text{base}}$.}
Since $G_{\text{base}}$ varies smoothly over the sphere, it is well-approximated by a low-order SH expansion:
\begin{equation}
G_{\text{base}}(\theta,\phi) \;\approx\; \sum_{l=0}^{L_{\text{base}}}\sum_{m=-l}^{l} c_{lm}\,Y_l^m(\theta,\phi),
\end{equation}
where $Y_l^m$ denotes the degree-$l$ order-$m$ spherical harmonic basis function, and $c_{lm}$ are expansion coefficients. The truncation order $L_{\text{base}}$ is small (e.g., $L_{\text{base}}=5$ in our experiments), yielding $(L_{\text{base}}+1)^2$ coefficients. In our implementation we use a real-valued SH basis (cos/sin form) to produce real coefficients with the same representational capacity as the complex SH basis.

\begin{algorithm}[t]
\caption{Prominence-Based Gaussian Center Selection}
\label{alg:center_select}
\begin{algorithmic}[1]
\State \textbf{Input:} residual grid $R(\theta_i,\phi_j)$, target count $K$
\State \textbf{Output:} centers $\{(\theta_k,\phi_k)\}_{k=1}^{K}$
\If{$\max(R)\le 0$}
    \State \textbf{return} $\{(\tfrac{\pi}{2},0)\}_{k=1}^{K}$ \hfill{\scriptsize\textit{(degenerate fallback)}}
\EndIf
\State Smooth residual: $R_\sigma \leftarrow \mathcal{G}_\sigma * R$ \hfill{\scriptsize $\sigma$: smoothing width}
\State Candidate peaks: $\mathcal{M}\leftarrow\{(i,j): R_\sigma(i,j)\ \text{is a local max}\}$
\State Threshold: $\mathcal{M}\leftarrow\{(i,j)\in\mathcal{M}: R_\sigma(i,j)\ge \tau\max(R_\sigma)\}$ \hfill{\scriptsize $\tau\in(0,1)$}
\For{each $(i,j)\in\mathcal{M}$}
    \State Prominence: $p_{ij} \leftarrow R(i,j) - \min\limits_{(i',j')\in\mathcal{W}_{ij}} R(i',j')$
\EndFor
\State Sort candidates by $p_{ij}$ (descending)
\State Initialize centers $\mathcal{C}\leftarrow\emptyset$
\For{candidates $(i,j)$ in sorted order}
    \If{$|\mathcal{C}|=K$} \textbf{break} \EndIf
    \If{$(i,j)$ is at least $\Delta$ away from all centers in $\mathcal{C}$}
        \State Add center $(\theta_i,\phi_j)$ to $\mathcal{C}$ \hfill{\scriptsize\textit{(non-max suppression)}}
    \EndIf
\EndFor
\If{$|\mathcal{C}|<K$}
    \State Fill remaining centers using the largest values of $R$ (greedy top-$K$ fallback)
\EndIf
\State \textbf{return} $\mathcal{C}$
\end{algorithmic}
\end{algorithm}

\textbf{Anisotropic Gaussians for $G_{\text{local}}$.}
Sharp lobes are spatially localized in $(\theta,\phi)$, motivating a sparse set of localized kernels:
\begin{equation}
G_{\text{local}}(\theta,\phi) \;\approx\; \sum_{k=1}^{K} a_k \,\mathcal{G}_k(\theta,\phi),
\end{equation}
where $K$ is the number of localized kernels and $a_k\ge 0$ is the amplitude of the $k$th kernel. We use an anisotropic Gaussian kernel in angular coordinates,
\begin{equation}
\mathcal{G}_k(\theta,\phi)
=\exp\!\left(
-\frac{1}{2}\Bigl(
\frac{(\theta-\theta_k)^2}{\sigma_{\theta,k}^2}
+\frac{d_\phi(\phi,\phi_k)^2}{\sigma_{\phi,k}^2}
\Bigr)\right),
\label{eq:gauss_kernel}
\end{equation}
where $(\theta_k,\phi_k)$ is the kernel center and $(\sigma_{\theta,k},\sigma_{\phi,k})$ are angular widths. We use the wrapped azimuth difference
\begin{equation}
d_\phi(\phi,\phi_k) \;=\; \mathrm{wrap}(\phi-\phi_k)\in(-\pi,\pi],
\end{equation}
which properly handles $2\pi$ periodicity, where $\mathrm{wrap}(\cdot)$ maps an angle to its equivalent value in $(-\pi,\pi]$. In the implementation used for our reported results, center coordinates $(\theta_k,\phi_k)$ are selected by a robust peak-finding procedure (Stage~2) and held fixed during refinement; the parameters optimized are $(a_k,\sigma_{\theta,k},\sigma_{\phi,k})$.

\subsubsection{Additive Model Rationale}
We use an additive decomposition as a practical and stable functional approximation that separates smooth structure from localized features. This form produces a well-conditioned fitting procedure. The global SH update is a linear least-squares problem, while the local Gaussian update is a bounded nonlinear least-squares problem. Empirically, this separation captures the dominant ``background + lobes'' structure of measured patterns while keeping the parameterization compact and interpretable.

\begin{figure*}[t]
    \centering
    \includegraphics[width=1\linewidth]{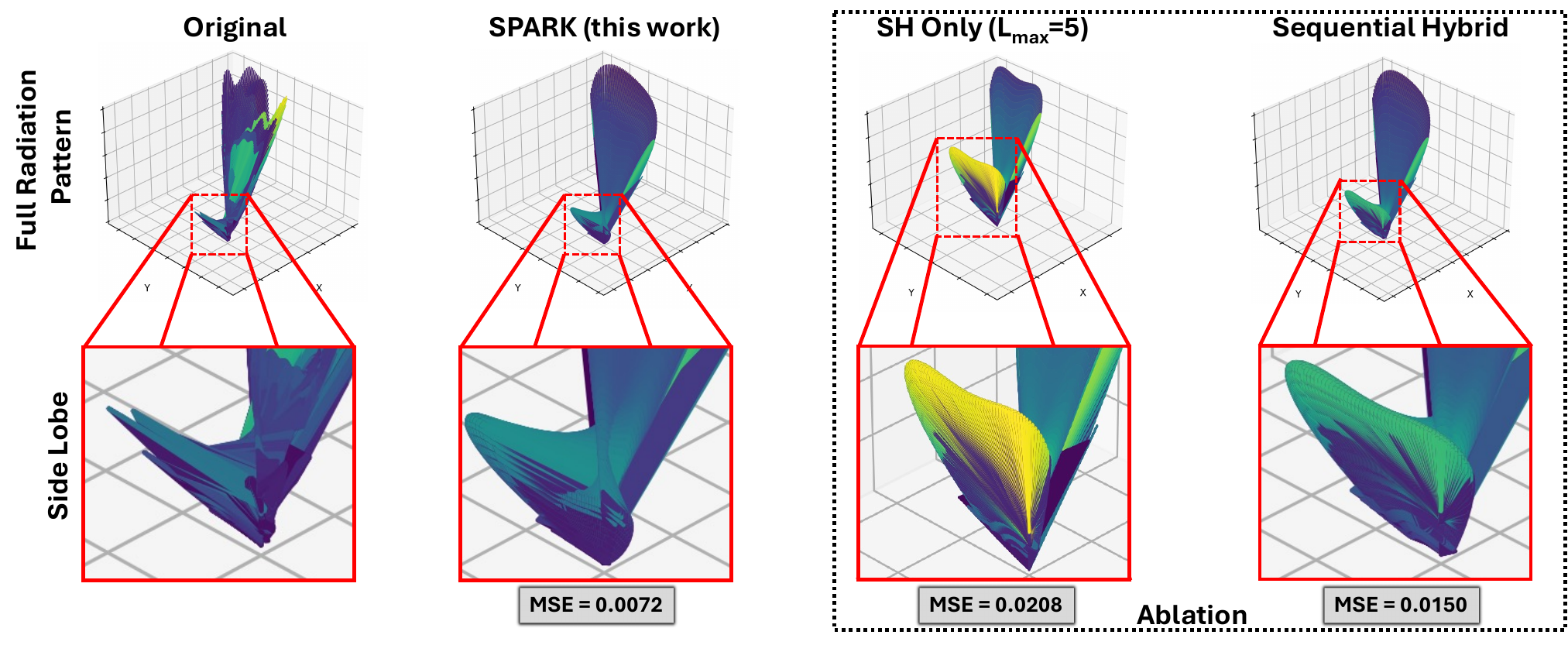}
    \caption{Visual comparison for the SA-1400-5900 antenna at 3.5~GHz. The top row shows original measured pattern, SPARK (joint refinement), SH-only ($L_{\max}{=}5$), and sequential hybrid (no joint refinement). The bottom row shows the magnified sidelobe region. SPARK better preserves fine sidelobe structure compared to the ablated models.}
    \label{fig:aerpaw_visual}
\end{figure*}

\subsection{Hybrid Fitting Methodology}
Given samples $\{g_{ij}=G(\theta_i,\phi_j)\}$ on a grid with $N=N_\theta N_\phi$ total samples, SPARK first fits the global SH component, then explains remaining structure using $K$ localized kernels, and finally refines both components by alternating updates.

\subsubsection{Stage 1: Spherical Harmonic Base Fit}
Let $\mathbf{g}=\mathrm{vec}(\{g_{ij}\})\in\mathbb{R}^{N}$ be the vectorized pattern samples. We form the SH design matrix $\mathbf{Y}\in\mathbb{R}^{N\times (L_{\text{base}}+1)^2}$ with entries
\begin{equation}
\mathbf{Y}_{n,(l,m)} \;=\; Y_l^m(\theta_n,\phi_n),
\end{equation}
and solve the least-squares problem
\begin{equation}
\hat{\mathbf{c}} \;=\; \arg\min_{\mathbf{c}}\;\|\mathbf{g}-\mathbf{Y}\mathbf{c}\|_2^2,
\end{equation}
yielding $\hat{G}_{\text{base}}(\theta,\phi)=\sum_{l,m}\hat{c}_{lm}Y_l^m(\theta,\phi)$.

\subsubsection{Stage 2: Residual and Prominence-Based Center Selection}
We compute a nonnegative residual that emphasizes lobe-like structure:
\begin{equation}
R(\theta,\phi) \;=\; \max\!\bigl(G(\theta,\phi)-\hat{G}_{\text{base}}(\theta,\phi),\,0\bigr).
\label{eq:residual_clip}
\end{equation}
Gaussian centers are selected from prominent peaks of $R$ using Algorithm~\ref{alg:center_select}. The goal is to capture distinct lobes rather than repeatedly selecting adjacent samples from the same lobe or selecting spurious measurement spikes. In our implementation, peak detection and separation respect azimuth periodicity (via wrapped $\phi$ differences), and non-maximum suppression enforces a minimum separation in $(\theta,\phi)$.

The prominence score $p_{ij}$ measures how much a peak rises above its local neighborhood minimum over a window $\mathcal{W}_{ij}$. This favors peaks that stand out from their surroundings and produces robust, well-separated centers. In our implementation, the separation test uses wrapped $\phi$ differences and an angular separation threshold $\Delta$ in $(\theta,\phi)$ coordinates.

\subsubsection{Stage 3: Bounded Nonlinear Least Squares for Gaussian Parameters}
With centers fixed, we fit Gaussian amplitudes and widths by minimizing
\begin{equation}
\hat{\boldsymbol{\psi}} \;=\;
\arg\min_{\boldsymbol{\psi}}
\sum_{i,j}\Bigl(R(\theta_i,\phi_j)-\sum_{k=1}^{K} a_k\,\mathcal{G}_k(\theta_i,\phi_j)\Bigr)^2,
\end{equation}
where
\begin{equation}
\boldsymbol{\psi}=[a_1,\ldots,a_K,\sigma_{\theta,1},\ldots,\sigma_{\theta,K},\sigma_{\phi,1},\ldots,\sigma_{\phi,K}]^T.
\end{equation}
We solve this as a bounded nonlinear least-squares problem using a trust-region reflective solver. We use bounds
\begin{align}
a_k &\in [0, 1.2],\\
\sigma_{\theta,k},\,\sigma_{\phi,k} &\in [0.01, 0.6]\ \text{radians}.
\end{align}
We initialize $a_k$ from the residual value near the selected center and initialize widths to a nominal value (e.g., $0.12$ radians). Note that although $G\in[0,1]$ after min--max normalization, the clipped residual can exceed 1 when the SH base undershoots (e.g., negative base values), motivating an amplitude upper bound slightly above 1.

\subsubsection{Stage 4: Joint Alternating Refinement}
Sequential fitting can leave coupling between the SH base and Gaussian components. We therefore refine the solution by alternating between (i) updating SH coefficients given the current Gaussian reconstruction and (ii) updating Gaussian amplitudes/widths given the current SH base. For a small number of iterations (e.g., 3--4 in our experiments), we perform:
\begin{enumerate}
    \item \textbf{Update SH coefficients:} with current local reconstruction $\hat{G}_{\text{local}}$, solve
    $\hat{\mathbf{c}}\leftarrow\arg\min_{\mathbf{c}}\|\mathbf{g}-\mathbf{Y}\mathbf{c}-\mathbf{g}_{\text{local}}\|_2^2$.
    \item \textbf{Update Gaussian parameters:} with updated base $\hat{G}_{\text{base}}$, recompute the clipped residual $R_{\text{new}}=\max(G-\hat{G}_{\text{base}},0)$ and solve a bounded nonlinear least-squares problem for $\boldsymbol{\psi}$ with fixed centers.
\end{enumerate}
This refinement is stable because the SH update is convex and the Gaussian update is bounded. In our experiments it reduces error substantially compared to a single-pass sequential fit.

\subsection{Pattern Reconstruction and Evaluation}
Given fitted parameters $\{\hat{c}_{lm}\}$ and $\{(\hat{a}_k,\hat{\theta}_k,\hat{\phi}_k,\hat{\sigma}_{\theta,k},\hat{\sigma}_{\phi,k})\}$, reconstruction is
\begin{equation}
\hat{G}(\theta,\phi) \;=\; \sum_{l=0}^{L_{\text{base}}}\sum_{m=-l}^{l}\hat{c}_{lm}Y_l^m(\theta,\phi)
\;+\;\sum_{k=1}^{K}\hat{a}_k\,\mathcal{G}_k(\theta,\phi).
\end{equation}
In our experiments we clip the reconstructed pattern to the normalized range, $\hat{G}\leftarrow \min(\max(\hat{G},0),1)$, which prevents negative values from the SH base and maintains consistency with the normalized input patterns. The per-direction evaluation cost is
$O(N_{\text{SH}}+K)$, where $N_{\text{SH}}=(L_{\text{base}}+1)^2$ is the number of SH basis functions, enabling fast downstream use after offline fitting.

\section{Results}
\label{sec:results}

We validate SPARK on two real-world measurement datasets and then connect the compression gains to a system-level impact model. We first evaluate SPARK on high-resolution 3D antenna radiation patterns measured in an anechoic chamber via the AERPAW testbed. We then show that the same global-plus-local decomposition adapts naturally to 1D RIS azimuth cuts. Finally, we quantify how reducing the payload required to represent hardware patterns can translate into uplink resource savings and goodput gains under a fixed uplink budget.

\subsection{Compression of 3D Antenna Patterns (AERPAW)}
\label{subsec:aerpaw_results}

\subsubsection{Dataset Description}
The AERPAW dataset provides high-resolution anechoic chamber measurements of commercial antennas~\cite{mma2-0t93-23}. We analyze three antenna types: two orientations of a wideband surface-mount monopole (RM-WB1-DN-BLK) and a triband stub antenna (SA-1400-5900). For each antenna, full 3D far-field patterns were measured. Each pattern contains 7,260 samples on a spherical grid.

All fitting and reconstruction errors in this section are computed in the min--max normalized linear domain defined in Eq.~\eqref{eq:minmax_norm}. This matches the domain used by our optimization procedure (linear least squares for the SH base and bounded nonlinear least squares for the Gaussian kernels). Unless otherwise noted, we set $L_{\text{base}}=5$ and $K=4$, i.e., 36 SH coefficients plus 12 Gaussian amplitude/width parameters (48 continuous parameters total, with centers chosen by Algorithm~\ref{alg:center_select}).

\subsubsection{Compression Performance}
Table~\ref{tab:aerpaw_mse_results} summarizes reconstruction MSE versus parameter count at four representative frequencies (915, 2100, 3500, 5500~MHz). Across antennas and bands where the pattern contains pronounced narrow lobes and structured sidelobes, SPARK improves fidelity per parameter over global-only SH representations. For example, for RM-WB1-DN-BLK (Upside Down) at 2.1~GHz, SPARK achieves $6.69\times10^{-3}$ MSE compared to $14.87\times10^{-3}$ for SH with $L{=}10$ (121 parameters), a $\approx2.22\times$ reduction while using fewer than half the parameters. This illustrates an insight: increasing SH order increases capacity globally, but can exhibit diminishing returns when the error is dominated by a small number of localized, high-curvature features.

For smoother patterns, or bands where the pattern is well explained by a low-order global envelope, SH can be competitive. In Table~\ref{tab:aerpaw_mse_results}, SA-1400 at 915~MHz is already well modeled by low-order SH, and SPARK provides only a modest improvement. Likewise, SA-1400 at 2.1~GHz shows that the hybrid model is not universally better. SPARK is comparable to the SH baselines but slightly worse than the higher-order SH setting at that frequency. This behavior is expected as SPARK allocates extra degrees of freedom specifically to localized features. When the underlying pattern is sufficiently smooth, the benefit of the local component can be limited.

Fig.~\ref{fig:aerpaw_visual} provides a qualitative view at 3.5~GHz. The SH-only model captures the broad directivity but smooths away fine sidelobe structure. The sequential hybrid fit improves detail relative to SH-only, but still leaves structured mismatch in sidelobes because the global and local components are fit in a single pass and can ``compete'' to explain the same energy. SPARK’s alternating refinement rebalances this split and better preserves sidelobes and null structure. While main-lobe accuracy is critical for link budget, sidelobe and null fidelity directly impacts interference leakage, spatial reuse, and the reliability of beam selection under multiuser scheduling. In that sense, SPARK’s improvements target pattern regions that can determine whether a beam causes unintended interference outside the intended direction.

\begin{table}[t]
\centering
\caption{MSE vs. parameter count (values scaled by $10^{-3}$).}
\label{tab:aerpaw_mse_results}
\renewcommand{\arraystretch}{1.15}
\setlength{\tabcolsep}{3.5pt}
\footnotesize
\begin{tabular}{@{}l c c c c@{}}
\toprule
\multirow{2}{*}{\textbf{Antenna}} & \multirow{2}{*}{\textbf{f (MHz)}} & \multicolumn{2}{c}{\textbf{SH MSE}} & \textbf{SPARK MSE} \\
\cmidrule(lr){3-4}
 &  & \textbf{36p} & \textbf{121p} & \textbf{48p} \\
\midrule
\multirow{4}{*}{RM-WB1 (Down)} 
 & 915  & 17.54 & 17.19 & \textbf{14.30} \\
 & 2100 & 15.10 & 14.87 & \textbf{6.69}  \\
 & 3500 & 18.84 & 18.99 & \textbf{10.14} \\
 & 5500 & 14.78 & 13.50 & \textbf{7.83}  \\
\midrule
\multirow{4}{*}{RM-WB1 (Up)}   
 & 915  & 19.45 & 19.06 & \textbf{12.92} \\
 & 2100 & 11.83 & 11.69 & \textbf{6.67}  \\
 & 3500 & 20.93 & 19.91 & \textbf{15.54} \\
 & 5500 & 15.36 & 14.64 & \textbf{10.21} \\
\midrule
\multirow{4}{*}{SA-1400}      
 & 915  & 4.83  & 4.05  & \textbf{4.59} \\
 & 2100 & 10.06 & 9.81  & \textbf{10.11} \\
 & 3500 & 20.83 & 20.73 & \textbf{7.15}  \\
 & 5500 & 26.08 & 25.60 & \textbf{15.26} \\
\bottomrule
\end{tabular}
\end{table}

\begin{figure*}[t]
    \centering
    \includegraphics[width=1\linewidth]{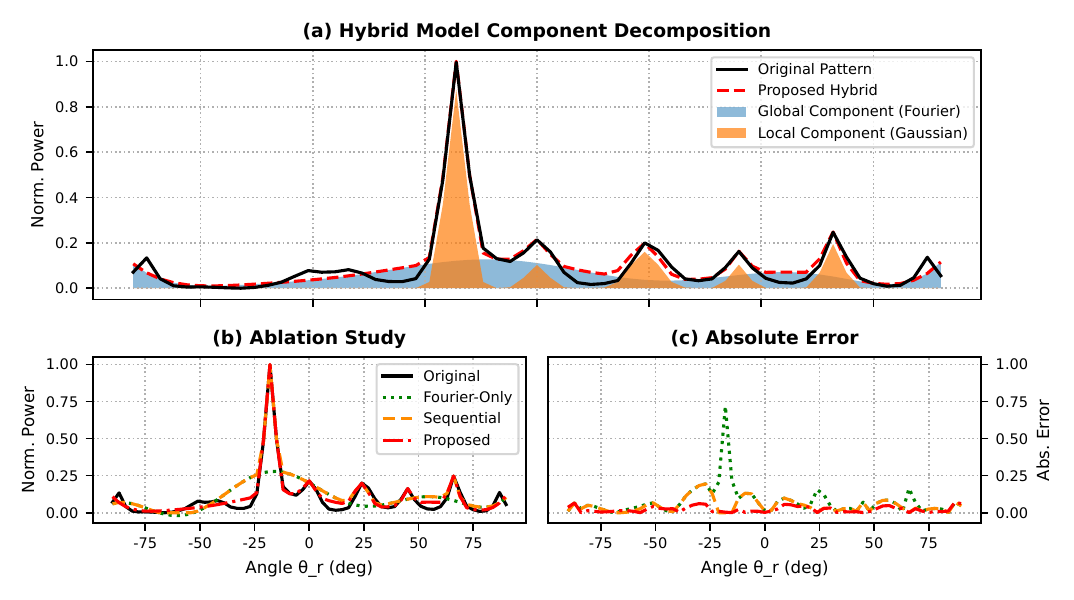}
    \caption{Analysis of SPARK on a representative RIS pattern ($\theta_t{=}20^\circ$, configuration $0^\circ;-30^\circ$). (a) Component decomposition showing the reconstruction is expressed as a smooth global Fourier component plus localized Gaussian components. (b) Ablation study comparing low-order Fourier-only, sequential hybrid (no alternating refinement), and the proposed alternating refinement. (c) Absolute error for each model, showing the error reduction achieved by alternating refinement.}
    \label{fig:ris}
\end{figure*}

\subsubsection{Storage and Transmission Implications}
SPARK reduces storage and transmission overhead by replacing dense pattern samples with a small set of parameters. A single AERPAW pattern contains 7,260 samples. Storing each sample as a 16-bit value requires 116,160 bits ($\approx$116~kbit). In contrast, under $L_{\text{base}}{=}5$ and $K{=}4$, SPARK uses 36 SH coefficients and 12 Gaussian amplitude/width parameters (48 continuous parameters). Using 16 bits for each SH coefficient and 12 bits for each Gaussian parameter produces 720 bits total, corresponding to a compression ratio of approximately 161:1.

\begin{table}[b]
\centering
\caption{Compression for a 7,260-sample AERPAW pattern under the SPARK setting $L_{\text{base}}=5, K=4$.}
\label{tab:compression_rate}
\renewcommand{\arraystretch}{1.15}
\setlength{\tabcolsep}{5pt}
\footnotesize
\begin{tabular}{@{}l c c c r@{}}
\toprule
\textbf{Representation} & \textbf{\# Values} & \textbf{Bits/Value} & \textbf{Total Bits} & \textbf{Ratio} \\
\midrule
Original samples & 7,260 & 16 & 116,160 & 1:1 \\
\midrule
\textbf{SPARK} & & & & \\
\quad SH coeffs ($L{=}5$) & 36 & 16 & 576 & \\
\quad Gaussians ($K{=}4$) & 12 & 12 & 144 & \\
\textbf{Total} & \textbf{48} & -- & \textbf{720} & \textbf{161:1} \\
\bottomrule
\end{tabular}
\end{table}

In our implementation, Gaussian centers $(\theta_k,\phi_k)$ are selected from the measurement grid by Algorithm~\ref{alg:center_select} and then held fixed during refinement. To transmit the model for reconstruction elsewhere, these centers can be encoded compactly as grid indices (e.g., $\lceil\log_2 N_\theta\rceil$ bits for $\theta$ and $\lceil\log_2 N_\phi\rceil$ bits for $\phi$ per Gaussian) or as quantized angles. This adds only a small overhead relative to the original 7,260-sample grid and reaffirms that SPARK converts a high-dimensional pattern into a small message that is practical to store, distribute, and reuse.

Beyond storage, the fitted representation is also inexpensive to use. After fitting, evaluating the pattern at any direction requires only $(L_{\text{base}}{+}1)^2$ SH basis evaluations plus $K$ Gaussian evaluations (Section~\ref{sec:model}). This supports downstream tasks that repeatedly query the pattern model. Tasks like scanning candidate directions, building high-resolution virtual codebooks, or scoring beams under different scheduling decisions can be completed without repeatedly handling dense pattern grids.

\subsubsection{Ablation: Impact of Joint Alternating Refinement}
Fig.~\ref{fig:aerpaw_visual} also serves as an ablation study. The SH-only model ($L_{\max}{=}5$) captures the broad directivity but fails to reproduce localized sidelobe structure. Adding Gaussians in a purely sequential fit improves fidelity, but can still leave structured mismatch because the global and local components are not rebalanced after the initial decomposition. SPARK’s alternating refinement resolves this coupling by repeatedly (i) refitting the SH base after local structure is explained and (ii) refitting the Gaussian parameters on the updated residual. This block-wise refinement is stable in our setting because the SH update is convex (linear least squares) and the Gaussian update is bounded (trust-region reflective nonlinear least squares). The result is a better allocation of model capacity. The SH base captures the smooth envelope while the Gaussians concentrate on sharp features that are inefficient to represent with a global basis alone.

\textbf{Takeaway.}
High-fidelity patterns show that most of the ``hard'' structure in real measured radiation patterns is not global. Instead, it is concentrated in a small number of sharp, localized lobes and sidelobes. SPARK captures the smooth envelope with a low-order SH base and spends a few additional parameters to model those localized features, avoiding the large parameter jump required by high-order SH expansions. This turns dense hardware fingerprints into compact, reusable descriptors that preserve the pattern details relevant to interference and beam decisions.

\subsection{Application to RIS Patterns}
\label{subsec:ris_results}

To demonstrate SPARK's versatility beyond 3D antenna patterns, we adapt it to measured patterns from a hardware RIS prototype~\cite{rossanese2022designing}. Each RIS measurement is a 1D azimuthal cut of received power versus receiver angle $\theta_r \in [-90^\circ,90^\circ]$.

\subsubsection{Dataset Description}
The dataset contains received power measurements of a $10{\times}10$ RIS board captured in an anechoic chamber. Measurements are provided for two fixed transmitter incidence angles ($\theta_t=20^\circ$ and $\theta_t=90^\circ$) across 1,891 programmed RIS configurations. For each configuration, received power is recorded over $\theta_r \in [-90^\circ,90^\circ]$. 

\subsubsection{Compression Performance}

We evaluate six representative patterns spanning boresight and off-axis steering ($\theta_n \in \{0^\circ,30^\circ,60^\circ\}$) for both $\theta_t=20^\circ$ and $\theta_t=90^\circ$ (with $\phi_n=-30^\circ$). As a strong parameter-matched baseline, we use a higher-order Fourier series with 10 harmonics (21 parameters). SPARK uses 4 Fourier harmonics plus $K=5$ Gaussians (19 parameters) and performs a small number of alternating refinement iterations that refit the Fourier coefficients and the Gaussian component on the updated residual.

Fig.~\ref{fig:ris} illustrates the need for SPARK, showing the behavior of the global and local modeling components. Fourier-only fits capture the broad envelope but can underfit sharp peaks, while SPARK better captures narrow structure using a small number of localized components.

Table~\ref{tab:ris_results} shows that SPARK reduces MSE in five out of six configurations while using fewer parameters. Improvement factors ranging from 1.6$\times$ to 10.4$\times$. The one case where the Fourier baseline is slightly better ($\theta_t=20^\circ$, $30^\circ;-30^\circ$) corresponds to a relatively smooth pattern for which a Fourier series alone is already effective. This leaves limited residual structure for localized kernels to explain.

\begin{table}[t]
\centering
\caption{RIS dataset: MSE improvement vs. Fourier baseline (MSE on normalized linear power).}
\label{tab:ris_results}
\renewcommand{\arraystretch}{1.10}
\setlength{\tabcolsep}{2.2pt}
\scriptsize
\begin{tabular}{@{}c c c c c c@{}}
\toprule
\textbf{$\theta_t$} & \textbf{RIS Config.} & \textbf{Fourier MSE} & \textbf{Seq. MSE} & \textbf{SPARK MSE} & \textbf{Improv.} \\
\midrule
$20^\circ$ & $0^\circ;-30^\circ$  & 0.00908 & 0.00438 & \textbf{0.00087} & \textbf{10.4$\times$} \\
$20^\circ$ & $30^\circ;-30^\circ$ & 0.00827 & 0.01946 & 0.00980 & 0.84$\times$ \\
$20^\circ$ & $60^\circ;-30^\circ$ & 0.00710 & 0.00835 & \textbf{0.00243} & \textbf{2.9$\times$} \\
\midrule
$90^\circ$ & $0^\circ;-30^\circ$  & 0.00844 & 0.00764 & \textbf{0.00257} & \textbf{3.3$\times$} \\
$90^\circ$ & $30^\circ;-30^\circ$ & 0.00785 & 0.00641 & \textbf{0.00216} & \textbf{3.6$\times$} \\
$90^\circ$ & $60^\circ;-30^\circ$ & 0.00921 & 0.01446 & \textbf{0.00564} & \textbf{1.6$\times$} \\
\bottomrule
\end{tabular}
\flushleft
\scriptsize{SPARK uses 19 parameters (4 Fourier harmonics + 5 Gaussians) vs.\ 21 for the Fourier baseline (10 harmonics). Seq.\ is the sequential hybrid fit (no alternating refinement).}
\end{table}

\subsubsection{Ablation: Alternating refinement is a driver of accuracy}
Fig.~\ref{fig:ris} explicitly isolates SPARK’s components. A low-order Fourier-only model underfits the sharp main lobe, while the sequential hybrid fit (single-pass Fourier + Gaussian on the residual) improves accuracy but still leaves structured error. Alternating refinement produces the largest gain by letting the global and local components re-balance. For the boresight case ($\theta_t=20^\circ$, $0^\circ;-30^\circ$), the sequential fit achieves $4.38\times10^{-3}$ MSE, while SPARK reduces this to $0.87\times10^{-3}$ (a $\approx5.0\times$ reduction over sequential), confirming that refinement is a core component of the method.

\subsubsection{Comparison with a Deep Learning Baseline}
To position SPARK against data-driven approaches, we train a CsiNet-like autoencoder \cite{wen2018deep} on all 1,891 RIS patterns to produce a compressed representation with the same dimensionality as SPARK (19 parameters). Table~\ref{tab:dl_comparison} shows the comparison results. The learned model achieves lower MSE on this dataset after training. However, it produces latent variables that are not physically interpretable, limiting its usefulness for semantic network control. In contrast, SPARK is training-free and produces actionable parameters (e.g., lobe widths and locations). This supports generalization to new hardware patterns without retraining and enables manipulation of the radiation pattern.

\textbf{Takeaway.}
RIS patterns show the same phenomenon as 3D antennas. A smooth global envelope plus a few sharp, configuration-dependent peaks explains most of the structure. SPARK leverages this by using a low-order Fourier base and a few Gaussians, achieving large accuracy gains over Fourier-only baselines at essentially the same parameter budget. The result is a compact, training-free RIS pattern description that preserves the features relevant to beam steering and interference behavior.

\subsection{System-Level Impact: CSI Overhead and Throughput Gain}

\begin{table}[t]
\centering
\caption{Comparison with a learned autoencoder baseline on a representative RIS pattern.}
\label{tab:dl_comparison}
\renewcommand{\arraystretch}{1.15}
\setlength{\tabcolsep}{4pt}
\footnotesize
\begin{tabularx}{\columnwidth}{@{}l c c c@{}}
\toprule
\textbf{Model} & \textbf{MSE ($\times 10^{-3}$)} & \textbf{Training} & \textbf{Interpretable?} \\
\midrule
Autoencoder & \textbf{0.16} & Required & No \\
SPARK (ours) & 0.87 & None & Yes \\
\bottomrule
\end{tabularx}
\end{table}

We now translate SPARK's pattern compression into a system-level networking impact by quantifying how reduced CSI feedback payload can increase uplink resources available for data. We run a Monte Carlo simulation ($10^5$ trials) that compares a simplified 5G NR-style codebook feedback baseline against a decoupled scheme enabled by SPARK. This simulation isolates feedback payload overhead within a fixed 10-slot uplink budget. Pilot/sounding resources and PHY-layer channel estimation are assumed equal across schemes and thus are not modeled.

\textbf{Baseline (5G NR-style) overhead model.}
Each user reports a CSI payload composed of a wideband component and subband components (PMI/CQI) across $N_{\text{sub}}{=}15$ subbands. In our configuration, this totals
$B_{\text{NR}} = 10 + 15\cdot(10+4) = 220$ bits per user per report, resulting in a per-cell overhead of $U\cdot B_{\text{NR}}$ bits for $U$ users.

\textbf{SPARK-enabled decoupled overhead model.}
The BS broadcasts the static antenna pattern model once as a 48-parameter message (768 bits using 16 bits/parameter), amortized over a coherence time of $T{=}100$ snapshots. In the uplink, each user reports a stochastic number of dominant environmental paths, modeled as $P_u = 1 + \mathrm{Poisson}(\lambda)$ with $\lambda{=}2$ (mean 3 paths/user). Each path is described by 3 parameters quantized to 16 bits each (48 bits/path). The resulting per-cell overhead is the amortized broadcast plus the sum of per-user path reports.

\begin{figure}[t]
\centerline{\includegraphics[width=\columnwidth]{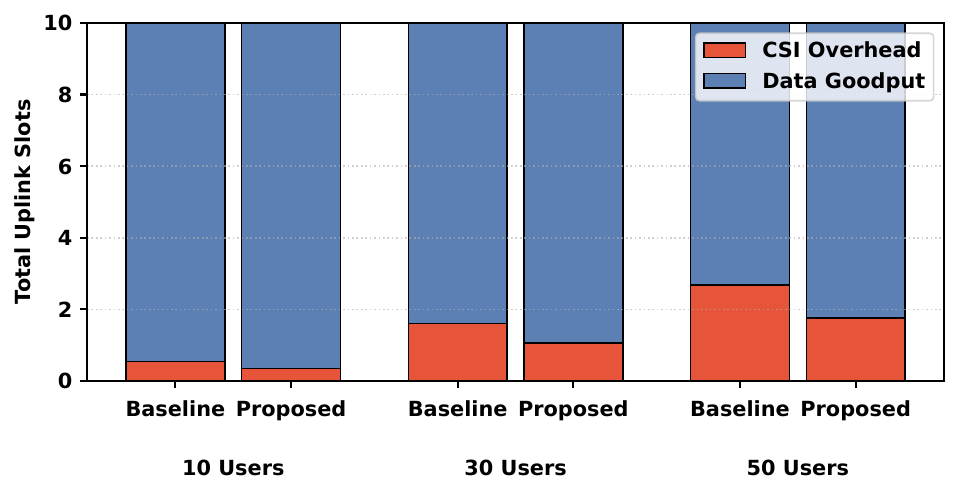}}
\caption{Uplink resource allocation showing the system-level benefit of SPARK. Stacked bars show the fraction of a fixed 10-slot uplink budget consumed by CSI feedback payload versus remaining data goodput for a 5G NR-style baseline and the SPARK-enabled scheme. At 50 users, SPARK reduces mean feedback from 11.00\,kb to 7.21\,kb (34.5\%), corresponding to 2.69 vs. 1.76 uplink slots of overhead (0.93 slots saved), which translates to a 12.65\% mean goodput gain.}
\label{fig:csi_overhead_plot}
\end{figure}

The results are visualized in Fig.~\ref{fig:csi_overhead_plot} and quantified in Table~\ref{tab:system_impact}. For a cell with 50 users, the SPARK-enabled scheme reduces mean uplink feedback overhead from 11.00 kb to 7.21 kb, freeing uplink resources for data. Under this fixed uplink budget, the saved overhead translates into a mean goodput gain of 12.65\% and an absolute throughput uplift of 13.88 Mbps in our fully loaded setting. The benefit increases with user count because the baseline feedback scales directly with $U$, while SPARK replaces a heavier per-user report with a compact path-based payload plus an amortized broadcast.

Lastly, we evaluate robustness to user mobility by sweeping the coherence time $T$, which controls how often the BS must amortize the broadcast of the static SPARK pattern model. The gain remains essentially unchanged across $T\in[10,500]$ (Table~\ref{tab:system_impact}, mobility footnote), varying by only 0.26 percentage points at 50 users. This occurs because the broadcast term is small relative to the per-user CSI payload, so SPARK's benefit is driven primarily by reducing the per-user feedback burden rather than relying on long coherence intervals.

\textbf{Takeaway.}
These results show that SPARK's pattern compression can translate into reclaimed uplink resources when feedback is the bottleneck. By amortizing a compact hardware description and shifting recurring reports toward sparse, environment-driven parameters, SPARK converts static pattern knowledge into persistent overhead savings that scale with user density.

\section{Limitations and Future Work}
\label{sec:lfw}

SPARK separates Gaussian center placement from subsequent parameter optimization. Centers are initialized heuristically using prominence-based peak selection on the residual and are held fixed during alternating refinement. This design produces stable and reproducible optimization, but it does not guarantee globally optimal lobe placement. An extension could include the center coordinates as optimization variables and jointly refine them with amplitudes and widths.

In our evaluation, each frequency is modeled independently. In practice, radiation patterns often evolve smoothly across adjacent frequencies. Additional compression could be possible by coupling parameters across frequency. One direction is to model the SH and Gaussian parameters as continuous functions of frequency via low-order polynomials or splines. Another extension could include sparsity-promoting penalties to automatically select the most salient lobes.

Our empirical results compare against classical baselines (SH/Fourier and ablations) and a learned autoencoder on RIS patterns. Broader comparisons to additional transforms such as wavelets or compressed sensing formulations could be valuable future work directions. In addition, our fitting and error metrics operate on per-pattern min--max normalized linear patterns, which preserves directional shape but removes absolute gain. Deployments that require calibrated gain could retain a scalar (e.g., peak/min/max) as side information or fit in a calibrated linear domain. Finally, our system-level simulation isolates feedback payload overhead to quantify potential uplink resource savings. It does not model pilot/sounding overhead, frequency-selective channels, or end-to-end estimation and scheduling in a live deployment. Integrating SPARK into a real-time feedback pipeline, like an O-RAN testbed, would enable evaluation under practical scheduler and PHY constraints.

\begin{table}[t]
\centering
\caption{System-level impact of SPARK vs. a 5G NR-style codebook scheme ($10^5$ trials, $T{=}100$).}
\label{tab:system_impact}
\renewcommand{\arraystretch}{1.2}
\setlength{\tabcolsep}{4pt}
\footnotesize
\begin{tabular}{@{}c cc cc@{}}
\toprule
\textbf{\# Users} & \multicolumn{2}{c}{\textbf{CSI Overhead (kb)}} & \textbf{Goodput Gain} & \textbf{Uplink Uplift} \\
\cmidrule(lr){2-3}
& \textbf{5G NR} & \textbf{SPARK} & \textbf{(\%)} & \textbf{(Mbps)} \\
\midrule
10 & 2.20 & \textbf{1.45 $\pm$ 0.21} & 1.94\% & 2.76 \\
30 & 6.60 & \textbf{4.33 $\pm$ 0.37} & 6.62\% & 8.32 \\
50 & 11.00 & \textbf{7.21 $\pm$ 0.48} & 12.65\% & 13.88 \\
\bottomrule
\end{tabular}

\vspace{0.5ex}
\flushleft
\footnotesize{\textit{Mobility sensitivity:} at 50 users, the mean goodput gain varies only from 12.42\% to 12.68\% when sweeping coherence time $T$ from 10 to 500 snapshots (0.26 percentage points).}
\end{table}

\section{Conclusion}
\label{sec:conclusion}

This paper introduced SPARK, a sparse hybrid representation for compressing antenna and RIS radiation patterns. SPARK decomposes patterns into a low-order global basis and sparse localized Gaussian kernels. It provides an interpretable and training-free representation that is efficient to store and evaluate. On measured 3D antenna patterns, SPARK achieves up to a 2.8$\times$ reduction in reconstruction MSE over higher-parameter SH baselines for sharp, directive patterns. Simultaneously, it compresses a 7,260-sample grid into a 48-parameter model. On a public RIS dataset of 1D azimuth cuts, SPARK achieves up to a 10.4$\times$ MSE reduction over a strong Fourier baseline while using fewer parameters. We further showed that these compression gains can translate into system-level benefits. In a simulation that isolates feedback payload overhead, a SPARK-enabled scheme results in a 12.65\% mean uplink goodput gain under a fixed uplink budget compared to a 5G NR-style codebook baseline. Overall, SPARK offers a practical step toward scalable, pattern-aware feedback and control by transforming dense measured pattern grids into compact, physically meaningful parameters. These parameters can be amortized and reused across network operations. The authors have provided public access to their code at \href{https://github.com/wbjorndahl/spark}{https://github.com/wbjorndahl/spark}.

\bibliographystyle{IEEEtran}
\bibliography{refs}

\end{document}